# Phase retrieval by pattern classification and circular mean for robust optical testing

Ohgan Kim[a], Bong Ju Lee[b], Yun-woo Lee[c, *], and Ho-soon Yang[c,d, **]

[a]Natural Science Research Institute, Korea Advanced Institute of Science and Technology, 291 Deahak-ro, Yusejong-gu, Deajon 34141, Republic of Korea
[b]Department of Advanced Green Energy and Environment, Handong Global University, 558 Handong-ro, Buk-gu, Pohang 37554, Republic of Korea
[c]Space Optics Team, Advanced Instrumentation Institute, Korea Research Institute of Standards and Science, 267 Gajeong-ro, Yuseong-gu, Daejeon 34113, Republic of Korea
[d]Science of Measurement, University of Science and Technology, 217 Gajeong-ro, Yuseong-gu, Daejeon 34113, Republic of Korea
*ywlee@kriss.re.kr, **hsy@kriss.re.kr

Abstract

For the optical testing of a large mirror with a long radius of curvature, it is generally necessary to use a single-shot phase-shifting interferometer to take several measurements because the influence of air turbulence on air stratification prevention should be reduced for an accurate measurement. In this paper, a new technique that applies hierarchical clustering, which classifies a few clusters according to the pattern similarity of acquired wrapped phases, is proposed. As the minority patterns of the wrapped phases are excluded, the effects of unknown noises can be reduced. For each cluster, the circular mean is used to calculate the denoised wrapped phases. The surface figure is obtained from the unwrapped phases. Because the running number of the phase-unwrapping algorithm becomes the same as the number of chosen clusters, the proposed technique is much faster than conventional procedures. In 500 measurements using a spherical mirror with a diameter of 1.5 m, the measurement time and standard deviation of the surface figure's root-mean-square error are reduced by up to approximately 23% and 15%, respectively, of the results obtained via the conventional procedure.

Keywords: optical testing, phase retrieval, interferogram pattern, hierarchical clustering, circular mean

1. Introduction

For a better high-resolution imaging of ground- or space-based telescopes, larger and more accurate aspheric mirrors are required [1]. The development of large mirrors needs a high level of material quality, fabrication, and testing technologies [2–5]. Particularly, the surface figure error measurement for large optics is always challenging under an optical workshop's environment due to vibration, air stratification, and air turbulence that degrade the optical image quality [6]. Because stratified air causes a wavefront aberration due to the refractive index difference, it should be randomly mixed using fans [7].

As a single-shot phase-shifting interferometer (PSI) is insensitive to vibration and turbulence, it is widely used to measure the root-mean-square error (RMSE) of surface figures [8]. According to the conventional image-processing procedure of a PSI, random noises can be reduced by averaging hundreds of repeated measurements.

Because the method does not work for the data set of noisy wrapped phases, an individual wrapped phase has to be repetitively unwrapped by as many as the number of data acquisition [9]. Therefore, the measurement usually takes several minutes, and unknown low-frequency noises and vibrations in the environment can contribute to the surface figure error [10].

Thus far, many sophisticated approaches have been proposed for reducing noises in individual wrapped phases. A spatial filter with a window of a certain size in a two-dimensional image (e.g., moving averaging filter, median filter, Gaussian filter, and sine/cosine averaging filter) was widely studied [11]. Qian applied a Fourier transform filter into a two-dimensional window to eliminate the frequency caused by noises [12]. Jiang proposed a technique to reduce noises by applying an adaptive filter to estimate the direction and curvature of interference fringes [13]. A deep learning model trained using simulation data was applied to remove noises in real measurement data [14]. In differential interferometric synthetic aperture radar, the circular normal distribution was used to compare the spatial correlation between master and slave interference fringes and to effectively reduce noises through the minimization of the weighted circular variance.

Recently, Kim *et al.* reported that a geometrical approach to the phase unwrapping, not the $2\pi$ modular arithmetic [15]. Based on this understanding, since wrapped phases mapped on a cylindrical surface can be considered true phases, the direct application of the circular mean is expected to reduce noises in mapped wrapped phases as long as they follow a circular normal distribution. However, because all the acquired wrapped phases do not have this property, we classified them into a few clusters that have a similar pattern with wrapped phases. Then, the denoised wrapped phases were calculated by the circular mean of each cluster, which can lead to a more accurate surface figure error. With this technique, the running number of the phase-unwrapping algorithm can be reduced as many as the number of clusters, and phases with less similarity can be filtered out by excluding in each cluster. This technique expects to enhance measurement speed and repeatability. We call this technique as phase retrieval by pattern classification and circular mean (PPC) in this paper.

We applied the PPC to measure the surface figure of a reference spherical mirror, whose diameter is 1.5 m and radius of curvature is 6 m. The PPC can reduce the deterioration of image quality that is caused by air turbulence using fans that prevent air stratification. The remainder of this paper is organized as follows: In Section 2, we discuss the conventional procedure problem by conducting repeated measurements. In Section 3, the hierarchical clustering for image classification and the circular mean are described for implementing the PPC. In Section 4, the surface RMSE and standard deviation (SD) obtained through the PPC and conventional procedure are compared.

2. Problem of conventional image-processing procedure

Fig. 1 shows a picture of the surface figure error measurement for a 1.5-m-diameter reference spherical mirror using the commercial single-shot PSI. This mirror can be used to measure a large flat mirror using the Ritchey–Common setup [16]. The interferometer(PhaseCam® 5030, 4D technology) is located approximately 6 m away from the mirror owing to its radius of curvature, which means that the surface measurement can be significantly affected by air turbulence, vibration, and stratified air.

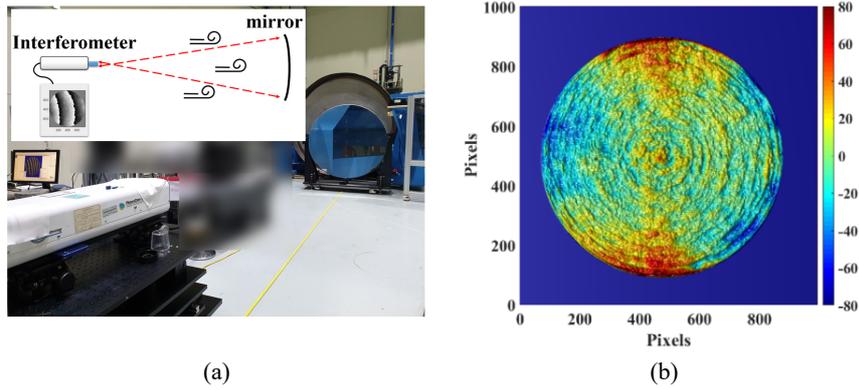

(a)             (b)

Fig. 1. Large mirror measurement with a single-shot PSI. (a) Picture and schematic drawing in this experiment; spiral icons stand for intended air turbulence to prevent stratified air. (b) One of the surface figures measured by 500 times. RMSE and SD are 22.2 nm and 2.6 nm, respectively.

Fig. 2 shows the schematic flow of the measurement in a commercial single-shot PSI [17]. The red box denotes the conventional image-processing procedure. The distorted wavefronts include the true phase ($\varphi$) of the mirror surface and noises ($\eta_i$) that are generated at each measurement ($i$-th measurement). The interferometer creates the wrapped phase ($\psi_i$) from the input distorted wavefront, conducts phase unwrapping, and averages unwrapped phases to provide the raw surface ($\tilde{\varphi}$). We can expect a reduction in the effect of random noises due to air turbulence or vibration by increasing the number of measurement ($N$) and averaging them to obtain results closer to the true phase.

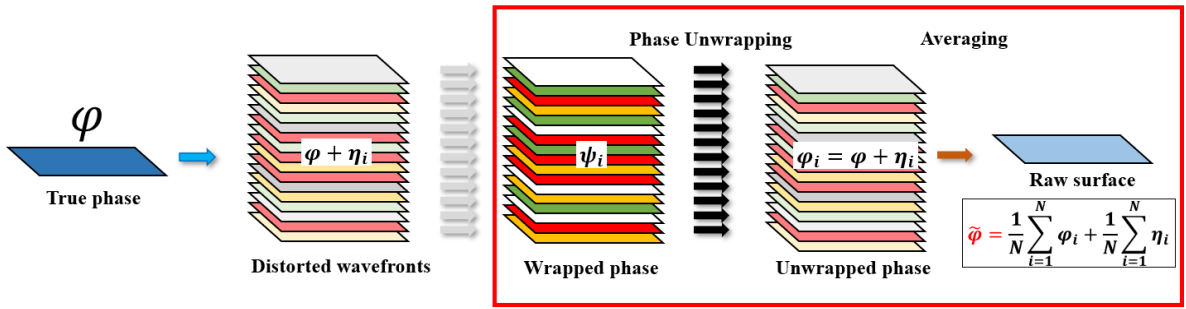

Fig. 2. Conventional image-processing procedure with $N$ measurements.

Fig. 3 (a) and (b) present the RMSE mean values and SD of the mirror surface figure with five repeated measurements, where each measurement consists of $N$ = 100–500. The fan-on and fan-off conditions denote whether the large fans stir the air around the mirror or not. The tilt and power terms are removed in the wavefront reconstruction. As shown in this figure, the surface RMSE mean value is 1–2.4 nm lower for the fan-on condition than for the fan-off. Thus, air turbulence generated by fans can reduce the aberration due to air stratification or non-uniform distribution of the refractive index of air. However, the RMSE mean value and SD did not converge as the number of measurements is increased from 100 to 500. Hence, the number of measurements does not guarantee a reduction in the random error.

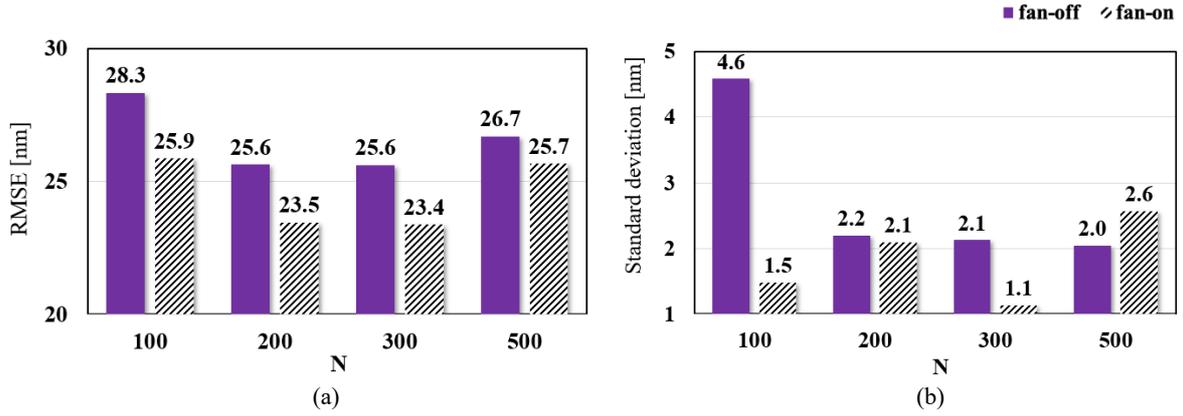

Fig. 3. Measurements using the conventional image-processing procedure versus *N* measurements.
(a) RMSE mean value of the surface figure, (b) SD of RMSE.

For *N* = 100, random noises were not reduced by the insufficient measurements. For *N* = 500, the SD was increased, which means that a large number of the measurement does not guarantee robustness of the results. These findings imply that a certain portion of measurements usually come from the low-frequency vibrations outside, even including unrecognizable unknown noises [10]. With more measurements, the number of events that generate unwanted data is random and their number is increased, so a technique for determining the bad data or not is important. However, a conventional interferometer cannot select the bad data from the repeated measurements, as shown in Fig. 2.

For better repeatability, a technique called smart averaging in the PSI's software drops out some surface figures depending on their RMSE values [18,19]. Because the running number of the phase-unwrapping algorithms is not reduced, the measurement still takes a long time and unknown noises are inevitable.

3. PPC: Phase retrieval by pattern classification and circular mean

Fig. 4 shows the schematic flow of the measurement with the PPC. The blue box shows the aspect of the data acquisition in the proposed technique. We assume that the true phase cannot be measured exactly, and the wrapped phases generated from the same observable status, which is called the perturbed phase ($\varphi_q$), have similar patterns. In addition, the distorted wavefronts measured from each perturbed phase have noise ($\eta_{q_i}$), which follows a circular normal distribution (*i* denotes the number of elements in the *q*-th perturbed phase).

The red box denotes the proposed image-processing technique. From randomly mixed wrapped phases, regardless of the number of clusters, the hierarchical clustering is suitable for classifying patterns among many clustering algorithms, such as hierarchical, partitioning, density-based, and model-based algorithms [20]. From each cluster, the circular mean calculates the denoised wrapped phases ($\overline{\psi}_q$). Because each perturbed phase ($\tilde{\varphi}_q$) can be reconstructed from them, the running number of the phase-unwrapping algorithm is the same as the number of classified clusters. The raw surface ($\tilde{\varphi}$) can be obtained by averaging the unwrapped perturbed phases.

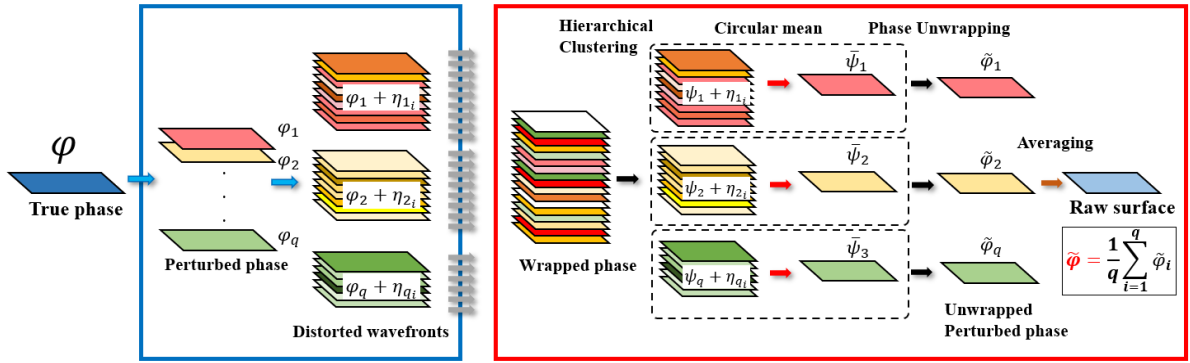

Fig. 4. Phase retrieval by pattern classification and circular mean.

3.1. Pre-processing for pattern classification

Fig. 5 shows the pre-processing for the hierarchical clustering. Each wrapped phase can be shifted as much as the piston aberration using Eq. (1), which is one of the identity equations of the $2\pi$ modular arithmetic. Here, the lower scripts "ac" and "ps" stand for the acquired wrapped phase and piston-shifted wrapped phase, respectively, and (i, j) denotes the center index pair of the wrapped phase. The pooling layer reduces the image size for a faster image classification while retaining its features and saving computing time and memory [21].

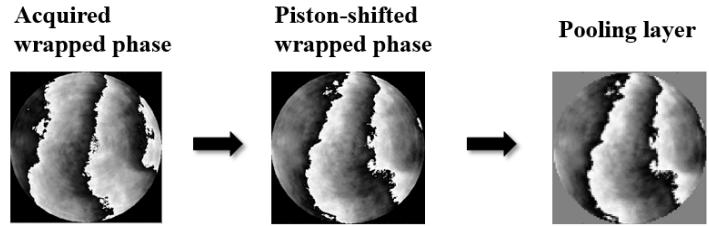

Fig. 5. Pre-processing for applying the hierarchical clustering.

$$\psi_{ps} = \psi_{ac} - \psi_{ac}(i, j) \ (\text{mod } 2\pi) \tag{1}$$

In this study, a $2 \times 2$ average pooling layer was used [22]. Because it reduces the image size to one quarter in one run, the repeated running of the pooling layer causes the loss of own features and reduces the accuracy of the pattern classification.

3.2. Hierarchical clustering based on the pattern of the wrapped phase

If there are pre-processed wrapped phases ($W_i$), as shown in Fig. 6 (a), the square root of the relative differences ($\Delta W_{ij} = \sqrt{(W_i - W_j)^2}$) is called the distance between every pair of objects. Fig. 6 (b) shows three relative difference of pairs and their distances. Using their distances, the relative position of each other is represented, as shown in Fig. 6 (c). Because W1 and W2 are relatively close, they can be classified as a cluster. The three wrapped phases can be classified into a cluster with two samplings and another cluster with one sample.

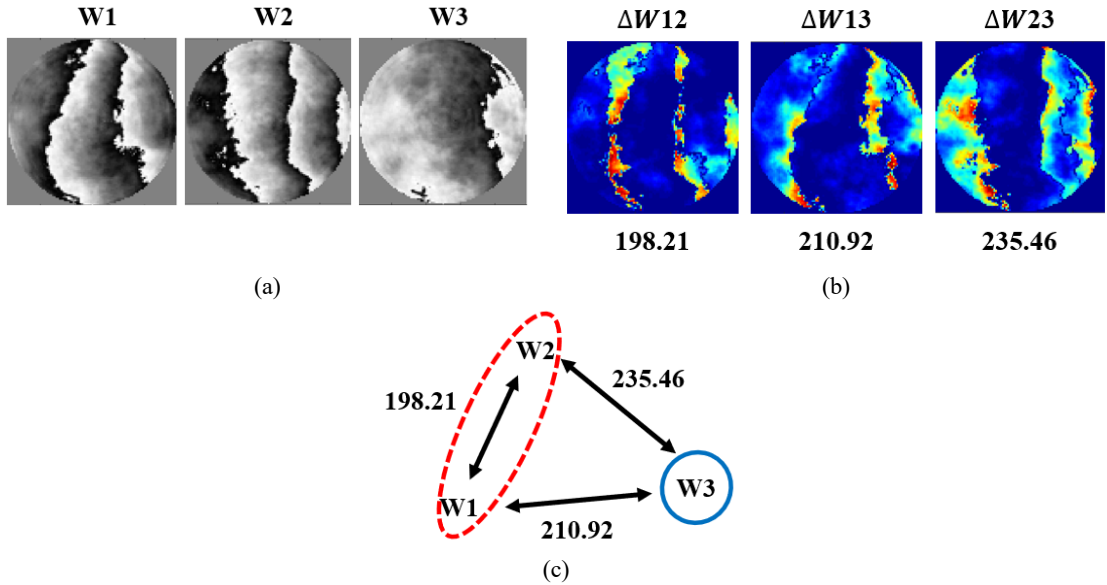

Fig. 6. Procedure for clustering. (a) Three different wrapped phases, (b) relative differences between every pair of wrapped phases, (c) positions of three wrapped phases by (b).

Fig. 7 (a) shows a dendrogram that is drawn based on every pair of the wrapped phases. The x-axis and y-axis denote the data acquisition order and normalized relative distance between measurements, respectively. By setting the minimum sampling number, we can find clusters that satisfy this condition. For example, if the minimum sampling number is assigned as 3, then six colored clusters (red, orange, yellow, green, blue, and purple) become chosen clusters and four black-colored clusters are abandoned due to insufficient sampling.

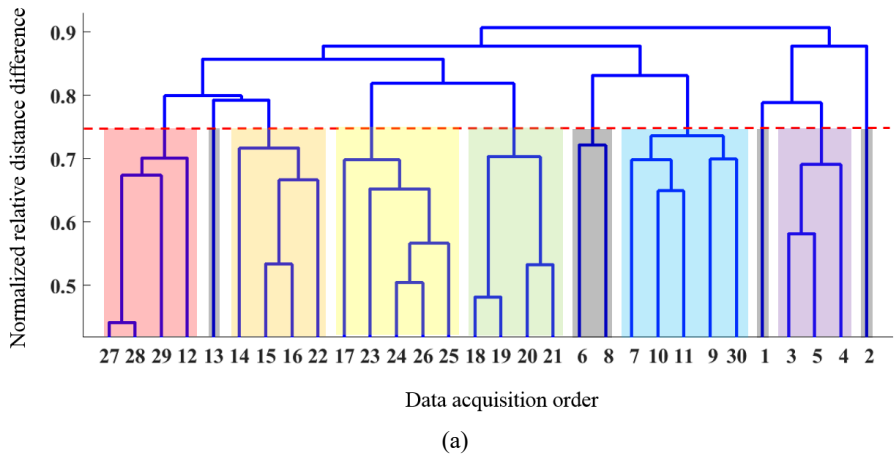

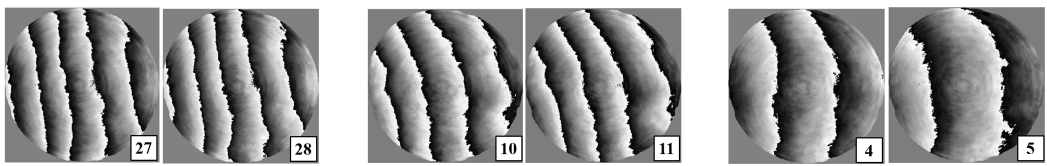

Fig. 7. Hierarchical clustering. (a) Dendrogram for image classification, (b) three examples of clustered wrapped phases (the number in the box denotes the data acquisition order).

In this technique, abandoned clusters can be considered contaminants due to unknown noises, so they are excluded. If the minimum sampling number is set to 1, then as the number of clusters is the same as the number of measurements, the conventional procedure is a special case of the proposed technique. Fig. 7 (b) represents three pairs in the chosen clusters.

3.3. Circular mean for the clustered wrapped phases

The wrapped phase value $\psi_i(m,n)$ corresponding to the arbitrary index $(m,n)$ of the *i*-th measured wrapped phase $\psi_i$ can be transformed into a point on the unit circle using Eq. (2).

$$x_i(m,n) = \cos(\psi_i(m,n)), y_i(m,n) = \sin(\psi_i(m,n)) \qquad (2)$$

The denoised wrapped value $\bar{\psi}(m,n)$ is obtained using Eq. (3). This value is the circular mean or the mean of circular quantities. It is statistically equivalent to the mean of the circular normal distribution or the von Mises distribution [23].

$$\bar{\psi}(m,n) = arctan2\left(\frac{1}{k}\sum_{i=1}^{k} y_i(m,n), \frac{1}{k}\sum_{i=1}^{k} x_i(m,n)\right) \qquad (3)$$

To identify the effect of the circular mean, the noisy surface was made by adding the surface of the peaks function, which is widely used in the validation of phase-unwrapping algorithms. The zero-mean Gaussian random noise was generated by the additive white Gaussian noise function. The noisy surface was wrapped by the atan2 function. They are the inbuilt functions of the MATLAB package [24]. We have generated eight wrapped phases with 512 x 512 pixels, peak-to-valley of 37.82 radians, and signal-to-noise ratio (SNR) of 5 dB.

Fig. 8 (a) shows a noisy wrapped phase and its residue positions detected by the branch cut algorithm of the Goldstein algorithm [25,26]. Fig. 8 (b) shows the averaged wrapped phase and its residue positions. Because a pair of residues prohibits path-following phase unwrapping in a certain area, many residues experience the phase-unwrapping failure. Fig. 8 (c) shows the wrapped phase by the circular mean and disappeared residues. A-A and B-B in Fig. 8 (b) and (c) stand for each 316th row line, respectively. Fig. 8 (d) shows the comparison of A-A and B-B with the 316th row line of the noise-free wrapped phase. The wrapped phase applied by the circular mean (red dashed line) shows the well-reconstructed height with respect to the noisy-free wrapped phase (blue dot line), but the averaged wrapped phase (black solid line) does not.

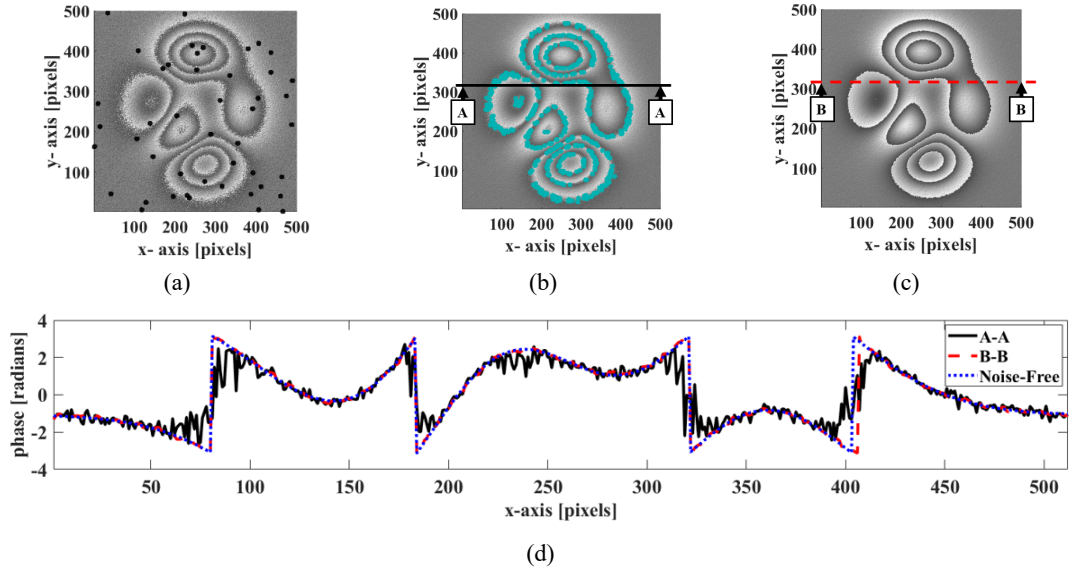

Fig. 8. Application of the circular mean for noisy wrapped phases of the peaks function surface. (a) One of the noisy wrapped phases (black spots are residues detected by the branch-and-cut algorithm), (b) averaged wrapped phase (cyan spots are residues) and its 316th row line (A-A), (c) denoised wrapped phase using the circular mean with no residue and its 316th row line (B-B), (d) comparison of A-A and B-B with the 316th row line of the noise-free wrapped phase.

4. Applications of the PPC

4.1. Data acquisition in a single-shot PSI

To apply the PPC, a commercial interferometer with a polarization pixelated camera was used to simultaneously obtain four phase-shifted interferograms, as shown in Fig. 9 (a). Each measurement was taken at 0.1 s intervals (exposure time of 1 ms). We used the wrapped phases calculated from the acquired phase-shifted interferograms, as shown in Fig. 9 (b). This experiment was repeated thrice in an environment with a fan-on condition.

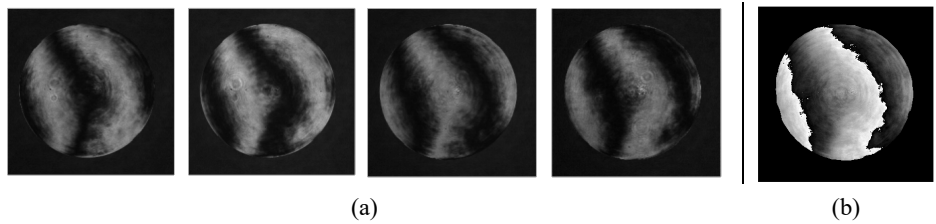

(a)                 (b)

Fig. 9. Data acquisition. (a) Four phase-shifted interferograms acquired from a single-shot PSI, (b) calculated wrapped phase using the four-bucket algorithm [17].

4.2. Application of the circular mean without pattern classification

Fig. 10 shows the RMSE of the surface figure obtained using the individual wrapped phase by the circular mean and the conventional image-processing procedure. It should be noted that the two latter image-processing procedures use accumulated wrapped phases from the first to the corresponding measurement. These procedures were performed in accordance with the operation of the fans. In Fig. 10 (a), the RMSEs of the conventional procedure and circular mean are very similar to each other and converge, but the individual unwrapped phase appears to be random. This finding implies that the RMSE can be obtained with the phase-unwrapping algorithm run just once, not 200 times. While measuring the surface figure error of a small mirror or a lens without a fan, a technique without a classification process may be applied.

Owing to the unknown cause that occurred after the 20th measurement in Fig. 10 (b), the RMSE of the circular mean becomes different from that of the conventional procedure. As the airflow and fans stabilize, the RMSE of the circular mean approaches that of the conventional procedure. However, in Fig. 10 (c), the results obtained via the circular mean exhibit irregular RMSE till the end. Thus, the noise distribution by the air turbulence and fans no longer follows the circular normal distribution. As mentioned earlier, because we assume that the similar patterns of wrapped phases have noise with the circular normal distribution, pattern classification must be preprocessed for applying the circular mean in the measurement of a large mirror.

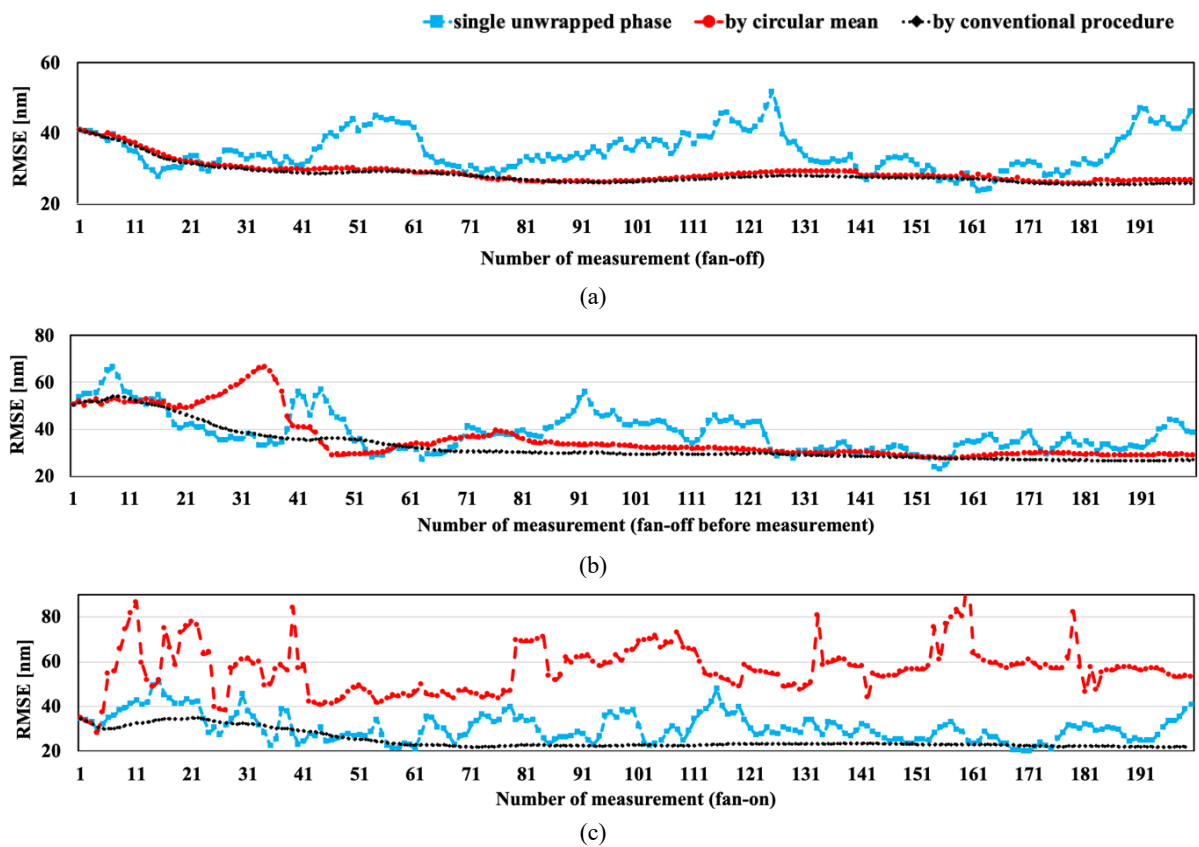

Fig. 10. Comparison of the surface figure RMSE obtained by the individual unwrapped phase, circular mean, and conventional procedure. (a) Fan-off, (b) fan-off before measurement, (c) fan-on during measurements.

To examine the experiment in Fig. 10 (c), a few wrapped phases and their averaged wrapped phase by the circular mean were shown in accordance with the order of measurements. In Fig. 11 (a), the tilt aberrations are changed, which is common in actual measurements. However, the patterns of their averaged wrapped phases cut and become weirdly connected, as shown in Fig. 11 (b). Because these patterns experience the failure of phase unwrapping, pattern classification of the wrapped phase is essential for applying the PPC in a real measurement environment. However, the complex patterns of the wrapped phase due to misalignments or rough surfaces is not suitable for the PPC.

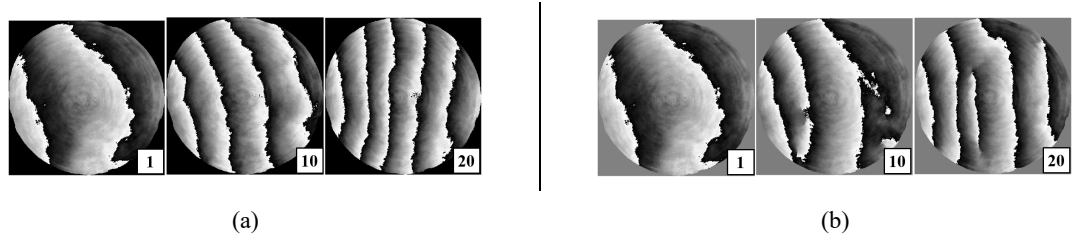

(a)　　　　　　　　　　　　　　　　　　　(b)

Fig. 11. Limitation of the circular mean due to various patterns. (a) 1st, 10th, and 20th wrapped phases in Fig. 10 (c), (b) patterns obtained by the circular mean (the number in the box denotes the number of measurement).

4.3. Results of the PPC

Fig. 12 shows the RMSE mean values and SD mean values of the results obtained by the PPC considering various SNRs for up to 500 repeated measurements compared with the results of the conventional procedure in Fig. 3. Because the SNR is the ratio of the number of useful data to contaminated data from unknown sources, each result of 16.9, 15.1, 13.8, 12.8, and 11.9 dB can be obtained by assigning the minimum sampling number from 2% up to 6% of the total number of measurements, which implies quite noisy situations [27].

The SD mean value of the surface figure RMSE was reduced by up to approximately 15%–50% of the results by the conventional procedure. In particular, when $N = 500$, although the conventional procedure seemed to increase the RMSE and SD mean values, the results of the PPC were both converged. Thus, the technique excluding insufficient sampling with respect to the similarity of the patterns can be a robust measurement in noisy environments. The more measurements you make, the more accurate results can be expected in the PPC.

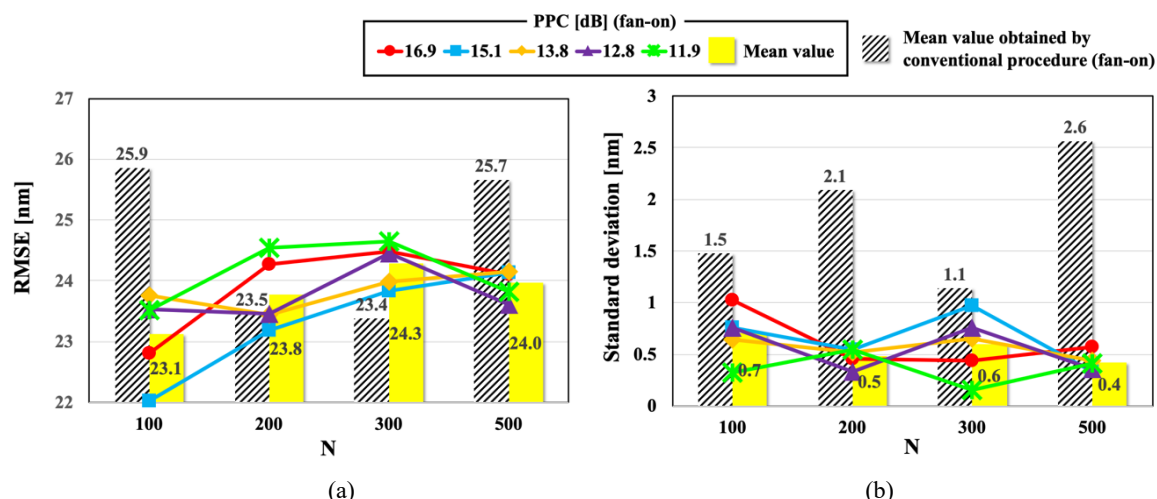

(a)　　　　　　　　　　　　　　　　　　　(b)

Fig. 12. Application of the PPC versus $N$ measurements in various SNRs. (a) RMSE, (b) SD of the RMSE of the surface figure. The minimum sampling numbers have SNRs of 16.9, 15.1, 13.8, 12.8, and 11.9 dB.

Table 1 summarizes the elapsed time with each detailed subroutine considering an SNR of 15.1 dB for one measurement. The ratio of the total elapsed time of the PPC to the total elapsed time of the conventional procedure decreases with the increase of the number of measurements because the running number of the phase-unwrapping algorithms does not depend on the number of measurements. If the PPC codes other languages and is optimized, then the process can be much faster. In this study, the used phase-unwrapping algorithm was developed in my previous work [15]. The PPC was performed in a personal computer with Intel i7 3.60 GHz processor, 16 GRAM memory, and MATLAB package (ver. 2018 b).

Table 1. Elapsed time of detailed subroutines according to the number of measurements

|   |   | Measurement number | | | |
|---|---|---|---|---|---|
|   |   | 100 | 200 | 300 | 500 |
|   | Subroutine list | Elapsed time [s] | | | |
| PPC | Data acquisition | 10 | 20 | 30 | 50 |
|   | Pooling layer | 1.0 | 2.1 | 3.1 | 5.6 |
|   | Hierarchical clustering | 1.5 | 6.2 | 11.2 | 0.2 |
|   | Circular mean | 0.9 | 1.5 | 1.8 | 3.4 |
|   | Phase unwrapping and Zernike poly. Fitting | 2.6 | 2.3 | 2.0 | 2.0 |
|   | Total measurement time | 21.8 | 44.4 | 55.3 | 72.7 |
| Conventional procedure | Measurement time | 63 | 126 | 189 | 315 |
|   | Time ratio (proposed technique/conventional procedure) | 36.3 % | 35.2 % | 29.2 % | 23.1 % |

Fig. 13 shows the classified patterns of the denoised wrapped phases and their raw phase for each experiment considering an SNR of 12.8 dB using 200 measurements, as shown in Fig. 12.

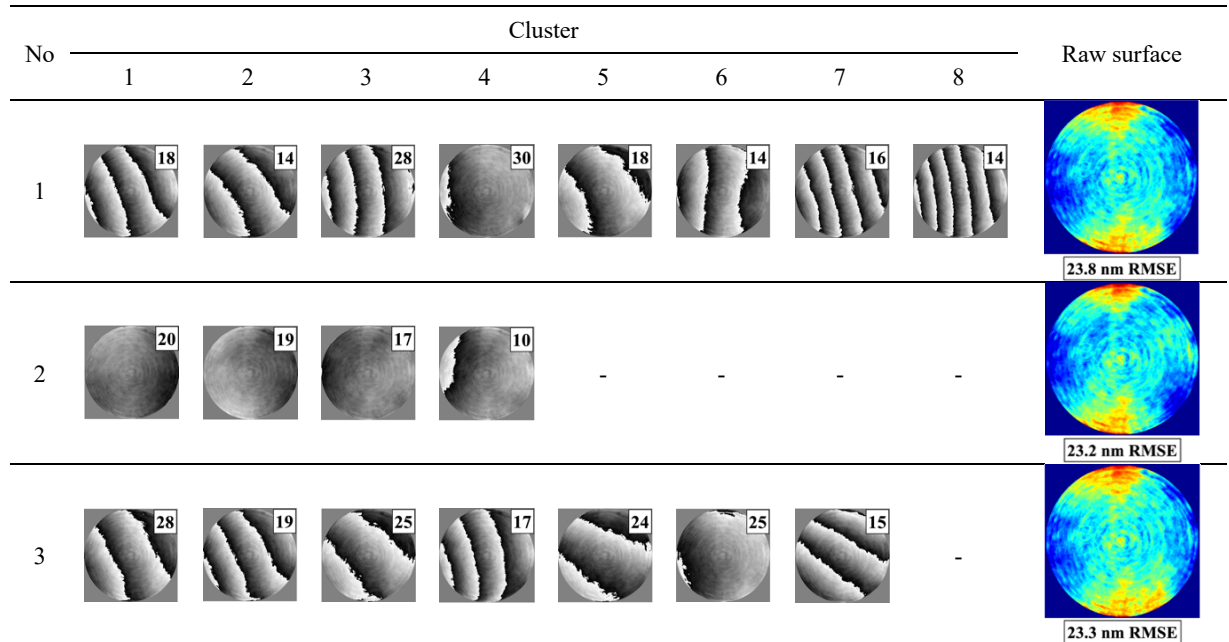

Fig. 13. Classified wrapped phases and raw surface for each experiment (the number in the box denotes the sampling number).

Fig. 14 shows some abandoned wrapped phases due to insufficient samplings in the third experiment presented in Fig. 13. In this technique, because these wrapped phases come from unknown noises, repeatability can be enhanced by excluding them.

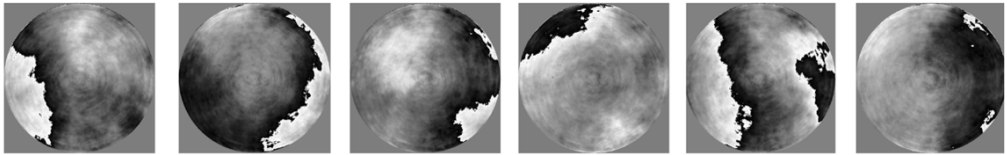

Fig. 14. Some abandoned wrapped phases in the third experiment presented in Fig. 13.

5. Conclusion

Here, we propose a new image-processing technique, called the PPC, to accurately measure large optics. In the conventional procedure, repeatability is worse with various low-frequency vibrations and unknown noises due to the long measurement time for measuring a 1.5-m-diameter reference spherical mirror. In the PPC, the acquired wrapped phases are classified into similar patterns using hierarchical clustering. By excluding insufficient sampling, their influences can be reduced. Moreover, the circular mean calculates the denoised wrapped phase for the chosen clusters. The surface figure error with better repeatability can be obtained by running the phase-unwrapping algorithm a few times. Therefore, the PPC provides a robust and fast measurement for the optical testing of large optics.

We expect that the PPC will enhance the productivity of polishing large mirrors and applicability related to various environments. The limitation of PPC is that when different types of patterns are included in the same cluster, the circular mean cannot be applied. In future work, if the performance of image classification and the application of the circular mean are enhanced, then faster and robust measurements can be expected.

Acknowledgements

This research was supported by a National Research Council of Science and Technology Grant funded by the Korean Government (CAP-12-04-KRISS).